\documentclass[11pt]{article}

\usepackage{amsfonts}
\usepackage{amsmath}
\usepackage{amssymb}
\usepackage{enumerate}
\usepackage{url}
\usepackage{graphicx}
\usepackage{sidecap}
\usepackage{lhelp}

\usepackage{bm}
\usepackage{color} 
\usepackage{multicol}
\usepackage{multirow}

\usepackage{amsthm}
\theoremstyle{definition}

\newcommand{\R}{\mathbb{R}}

\newcommand{\scad}{sub-CAD}
\newcommand{\scads}{sub-CADs}

\usepackage[boxed,vlined,ruled,linesnumbered]{algorithm2e}

\usepackage{fancyhdr}

\usepackage[
top    = 3.00cm,
bottom = 3.00cm,
left   = 3.00cm,
right  = 3.00cm]{geometry}

\title{\textbf{An implementation of Sub-CAD in Maple}}
\author{Matthew England and David Wilson}
\date{Department of Computer Science, University of Bath, Bath, UK \\ 
\texttt{M.England@bath.ac.uk} \qquad \texttt{D.J.Wilson@bath.ac.uk}}

\begin{document}

\maketitle

\pagestyle{fancy}
\lhead{M.~England and D.~Wilson}
\rhead{An implementation of Sub-CAD in Maple}

\begin{abstract}
Cylindrical algebraic decomposition (CAD) is an important tool for the investigation of semi-algebraic sets, with applications in algebraic geometry and beyond.  We have previously reported on an implementation of CAD in \textsc{Maple} which offers the original projection and lifting algorithm of Collins along with subsequent improvements.

Here we report on new functionality: specifically the ability to build cylindrical algebraic sub-decompositions (sub-CADs) where only certain cells are returned.  We have implemented algorithms to return cells of a prescribed dimensions or higher (layered {\scad}s), and an algorithm to return only those cells on which given polynomials are zero (variety {\scad}s).  These offer substantial savings in output size and computation time.

The code described and an introductory \textsc{Maple} worksheet / pdf demonstrating the full functionality of the package should accompany this report.
\end{abstract}

\noindent This work is supported by EPSRC grant EP/J003247/1.

\section{Introduction} 
\label{SEC:Intro}

This report concerns \texttt{ProjectionCAD}: a \textsc{Maple} package for cylindrical algebraic decomposition (CAD)  developed at the University of Bath.  The extended abstract \cite{EWBD14} at ICMS 2014 describes how this package utilises recent CAD work in the \texttt{RegularChains} Library of \textsc{Maple}, while still following the classical projection and lifting framework for CAD construction.  The present report is to accompany the release of \texttt{ProjectionCAD} version 3, describing the new functionality this introduced.  The report should be accompanied by the code described and an introductory \textsc{Maple} worksheet / pdf demonstrating the full functionality of the package. The previous two versions of \texttt{ProjectionCAD} are hosted alongside similar reports documenting their functionality \cite{England13a, England13b}.

Version 3 introduces functionality for \textbf{cylindrical algebraic sub-decompositions} ({\scad}s): subsets of CADs sufficient to describe the solutions of a given formulae.  Two distinct types are provided, whose theory was developed in \cite{WBDE14}. The first type contains only those cells of a certain dimension and higher, reducing both the output size and computational time by giving only output of the required generality.  We have implemented both a direct and recursive algorithm to build these \emph{layered {\scad}s}.  
The second type contains only those cells on which given equations are satisfied (lie on a prescribed variety).  When building a CAD for a formula with an equational constraint then only these cells can contain the solution set.  These \emph{variety {\scad}s} clearly reduce the output, and can also reduce computation time depending on the rank of the variety relative to the variable ordering. 

We continue the introduction by reminding the reader of the necessary background material on CAD and summarising our previous work with \texttt{ProjectionCAD}.  The following sections then describe the new functionality in \texttt{ProjectionCAD} version 3.

\subsection{Background on CAD}
\label{SubSEC:BkGd}

A \textbf{cylindrical algebraic decomposition} (CAD) is: a \textbf{decomposition} of $\R^n$ meaning a collection of non-intersecting \textbf{cells} whose union is $\R^n$; \textbf{algebraic} meaning each cell can be described by a finite sequence of polynomial relations; \textbf{cylindrical} meaning that with respect to a given variable ordering, the projection of any two cells onto a lower dimensional space (with respect to the ordering) is either equal or disjoint.  
CAD was introduced by Collins \cite{Collins1975} who provided an algorithm which given a set of polynomials would produce a CAD which was \textbf{sign-invariant}: so each polynomial had constant sign in each cell.  Collins developed CAD as a tool for quantifier elimination in real closed fields but CAD has since found applications as diverse as robot motion planning \cite[etc]{WDEB13} and algebraic simplification \cite[etc.]{DBEW12}.

Collins' algorithm (see for example \cite{ACM84I}) has two phases.  The first, \textbf{projection}, applies a \textbf{projection operator} repeatedly to a set of polynomials, each time producing another set in one fewer variables.  Together these contain the \textbf{projection polynomials}.  The second phase, \textbf{lifting}, builds the CAD incrementally.  First $\R$ is decomposed into cells which are points and intervals corresponding to the real roots of the univariate polynomials.  Then $\R^2$ is decomposed by repeating the process over each cell using the bivariate polynomials at a sample point of the cell.  The output for each cell consists of \textbf{sections} of polynomials (where a polynomial vanishes) and \textbf{sectors} (the regions between these). Together these form the  \textbf{stack} over the cell, and taking the union of these stacks gives the CAD of $\R^2$.  This process is repeated until a CAD of $\R^n$ is produced.  
Collins' original projection operator was defined so that the CAD of $\R^n$ produced using sample points in this way could be concluded sign-invariance. 
The key tool to conclude this was polynomials being \textbf{delineable} in a cell, meaning the zero set of individual polynomial are disjoint sections and the zero set of the sections from different polynomials are identical or disjoint.  More efficient projection operators have since been developed to conclude both sign-invariance and other invariance conditions.

The output of a CAD algorithm depends on the ordering of the variables.  In this paper we usually work with ordered variables $\bm{x} = x_1 \prec \ldots \prec x_n$, (so we first project with respect to $x_n$ and continue until we have univariate polynomials in $x_1$).  The \textit{main variable} of a polynomial, ${\rm mvar}(f)$, is the greatest variable present with respect to the ordering.  

All cells are equipped with a \textbf{sample point} and a \textbf{cell index}.  The index is an $n$-tuple of positive integers that corresponds to the location of the cell relative to the rest of the CAD. Cells are numbered in each stack during the lifting stage (from negative to positive), with sectors having odd numbers and sections having even numbers.  Therefore the dimension of a given cell can be easily determined from its index: simply the number of odd indices in the $n$-tuple. Note that the {\scads} we discuss later are \textbf{index-consistent}, meaning a cell in a {\scad} will have the same index as if it was produced in the associated complete CAD. 

Since Collins published the original algorithm there has been much research into improvements with a summary of the developments over the first twenty years given by \cite{Collins1998}.  Key developments since then include the use of certified numerics \cite{Strzebonski2006, IYAY09} in the lifting phase, projection operators for studying multiple formulae \cite{BDEMW13} and building CADs by first decomposing complex space and refining to real space, instead of projecting and lifting \cite{CMXY09, CM12b, BCDEMW14, EBCDMW14}.

\subsection{The \texttt{ProjectionCAD} package}

\texttt{ProjectionCAD} is a \textsc{Maple} package developed at The University of Bath to implement CAD via projection and lifting.  Its name was chosen to distinguish it from the CAD algorithm in \cite{CMXY09} which is distributed with \textsc{Maple} as part of the \texttt{RegularChains} library \cite[etc.]{MorenoMaza1999, LMX05}.  Nevertheless, the package makes much use of procedures from the \texttt{RegularChains} Library and the motivation and details for this are given in \cite{EWBD14}.

In \cite{England13a} we described our implementation of both McCallum's and Collin's algorithms to produce sign-invariant CADs.  In particular, we highlighted that this was the only implementation to offer order-invariant CADs (meaning each polynomial has constant order of vanishing on each cell) and delineating polynomials \cite{McCallum1998, Brown2005a} which modify the lifting phase to allow McCallum's projection operator to be applied more widely.  The latter meant that some examples of unnecessary failure in \textsc{Qepcad} \cite{Brown2003b} could be avoided, 
while the former was necessary for the extensions in the second release.  
These were described in \cite{England13b} and allowed for CADs invariant with respect to an equational constraint or CADs which were truth-table invariant (TTICADs) with respect to a list of formulae (meaning each formulae has constant Boolean truth value on each cell).  
The TTICAD theory was developed in \cite{BDEMW13, BDEMW15} and is based on an reduced projection operator analogous to McCallum's reduced operator for equational constraints \cite{McCallum1999b}.  In addition it was noted that the reduced projection theory allows also for improvement to the lifting phase, leading to \texttt{ProjectionCAD} offering lower cell counts than \textsc{Qepcad} even for examples with only one equational constraint.  The second release of the code also included heuristics to help with choices of problem formulation following \cite{DSS04, BDEW13}.

\section{Layered sub-CADs} 
\label{SEC:LCAD}

Define cells in a CAD with the same dimension as a {\bf layer} and let $\ell$ be an integer, $1 \leq \ell \leq n+1$.
Then an {\bf $\bm{\ell}$-layered sub-CAD} is the subset of a CAD of $\mathbb{R}^n$ with cells of dimension $n-i$ for $0 \leq i < \ell$.    They were introduced by the present authors in \cite{WBDE14} (although there had been previous work where only cells of full-dimension were computed \cite{McCallum1993, Strzebonski2000}).

Algorithm \ref{alg:layeredCAD} describes how an $\ell$-layered sub-CAD may be produced with the main idea to run the projection phase as normal (step \ref{step:P}), but then truncate in the lifting phase when cells of a high enough dimension cannot be produced.
The algorithm can run with different projection operators: that of Collins \cite{ACM84I} or McCallum \cite{McCallum1998} to build sign-invariant CADs for polynomials (as detailed in \cite{England13a}); or operators to build truth-invariant CADs for formulae using equational constraints \cite{McCallum1999b} and truth-table invariance \cite{BDEMW13, BDEMW15} (as detailed in \cite{England13b}).

We build the CAD of the real line as normal (step \ref{step:base}) but then for each successive lift we first check the dimension of the cell to be lifted over (step \ref{step:dimcheck}) and only proceed to generate the stack if cells of dimension $\ell$ or higher in $\mathbb{R}^n$ could be produced from it. The stack generation command is detailed in \cite{EWBD14} and makes use of the \texttt{RegularChains} stack generation code requiring a preprocessing step.  Some projection operators can incur theoretical failure for input that is not well-oriented, in which case this is identified during stack generation and FAIL returned.  A final loop is used (steps \ref{step:finalA} $-$ \ref{step:finalB}) to remove any lower dimensional cells that were produced as part of a stack.  The approach was verified in \cite{WBDE14}.

\begin{algorithm}[t]
\SetKwInOut{Input}{Input}\SetKwInOut{Output}{Output}

\Input{A choice of projection operator \texttt{ProjOp}; input $F(\bm{x})$ (polynomials or formulae) in ordered variables $\bm{x}$; and an integer $\ell \in (1,n+1)$.
}

\Output{An $\ell$-layered {\scad} for $F$, or FAIL if $F$ not well-oriented.}
\BlankLine
Run the projection phase and set ${\bf P}[i]$ to be the projection polynomials with mvar $x_i$\label{step:P}\;
Set $\mathcal{D}[1]$ to be the CAD of $\mathbb{R}^1$ obtained by isolating the roots of  ${\bf P}[1]$\label{step:base}\;
\For{$i=2,\ldots,n$}{
  Initialise $\mathcal{D}[i] := [ \ ]$\;
  \For{$c \in \mathcal{D}[i-1]$}{
    Evaluate ${\tt dim} := \sum_{\alpha \in c.{\tt index}} \left(\alpha \mod 2\right)$\label{step:dimcheck}\;
    \If{${\tt dim} > i-\ell-1 $ }{
    	$S := {\tt GenerateStack}({\bf P}[i],c)$\;
    	\eIf{$S=${\rm FAIL}}
    		{\Return {\rm FAIL}\tcp*{Input is not well oriented}}
    		{ Add the cells in $S$ to $\mathcal{D}[i]$\;}
     }
  }
}
Initialise $\mathcal{D} := [ \ ]$\;
\For{$c \in \mathcal{D}[n]$\label{step:finalA}}{
    Evaluate ${\tt dim} := \sum_{\alpha \in c.{\tt index}} \left(\alpha \mod 2\right)$\;
    \If{${\tt dim} > n-\ell$ }{
       Add $c$ to $\mathcal{D}$\label{step:finalB}\;
     }
}
\Return $\mathcal{D}$\;
\caption{${\tt LCAD}(F,\bm{x},\ell,\texttt{ProjOp})$: Algorithm to produce $\ell$-layered {\scad}s.
}
\label{alg:layeredCAD}
\end{algorithm}

As previously described in \cite{England13a} there are various output formats available.  At the least each cell is represented by an index (positioning it in the CAD) and a sample point (encoded as a regular chain and bounding box) while an algebraic description is also available.  Of particular note it the intuitive piecewise construction (described in \cite{CDMMXXX09}) which highlights the tree-like nature of CAD.  This has been adapted in \texttt{ProjectionCAD} to display layered sub-CADs with the truncated branches clearly indicated.

A simple example of using \texttt{LCAD} in \textsc{ProjectionCAD} now follows, appearing as it would in a {\sc Maple} worksheet, except that the sample points have been replaced by $SP$ for brevity.  
\begin{verbatim}
> LCAD([x^2+y^2-1], 1, [y,x], method=McCallum, output=piecewise);
\end{verbatim}
\[
\begin{cases}
SP & \quad x<-1 \\
*** & \quad \mbox{branch}=\mbox{truncated} \\
\begin{cases} 
SP & \qquad\qquad y<-\sqrt{-{x}^{2}+1} \\ 
*** & \quad \mbox{branch}=\mbox{truncated} \\
SP & -\sqrt {-{x}^{2}+1}<y<\sqrt {-{x}^{2}+1}  \\ 
*** & \quad \mbox{branch}=\mbox{truncated} \\
SP & \qquad\qquad \sqrt {-{x}^{2}+1}<y 
\end{cases}
&-1<x<1 \\
*** & \quad \mbox{branch}=\mbox{truncated} \\
SP & \quad 1<x
\end{cases}
\]

Here a sign-invariant sub-CAD for the defining polynomial of the unit circle has been produced.  The number $1$ as the second input indicates that only one layer of cells is to be produced: those five cells with full dimension.  Two simplifications have been in this sub-CAD compared to a complete CAD.  First when lifting over the CAD of the real line the two points ($x=\pm 1$) have not been lifted over: they failed the dimension check at step \ref{step:dimcheck} and were then excluded from the output during steps \ref{step:finalA} $-$ \ref{step:finalB}.  Second, those cells in the stack over the interval $(-1,1)$ on the real line withut full dimensional were also discarded in the final loop.  Note that while the second simplification reduced the output, the first also reduced the computation time since no real root isolation was required over 2 of the 5 cells in $\mathbb{R}^1$.

We could have instead asked for two layers of cells to receive the output below.
\begin{verbatim}
> LCAD([x^2+y^2-1], 2, [y,x], method=McCallum, output=piecewise);
\end{verbatim}
\[
\begin{cases}
SP & \quad x<-1 \\
\begin{cases} 
SP & \quad y<0 \\ 
*** & \quad \mbox{branch}=\mbox{truncated} \\ 
SP & \quad 0<y 
\end{cases}  
& \quad x=-1 \\
\begin{cases} 
SP & \qquad\qquad y<-\sqrt{-{x}^{2}+1} \\ 
SP & \qquad\qquad y=-\sqrt {-{x}^{2}+1} \\ 
SP & -\sqrt {-{x}^{2}+1}<y<\sqrt {-{x}^{2}+1}  \\ 
SP & \qquad\qquad y=+\sqrt {-{x}^{2}+1} \\ 
SP & \qquad\qquad \sqrt {-{x}^{2}+1}<y 
\end{cases}
&-1<x<1 \\
\begin{cases} 
SP & \quad y<0 \\ 
*** & \quad \mbox{branch}=\mbox{truncated} \\ 
SP & \quad 0<y 
\end{cases}  
& \quad x=1  \\
SP & \quad 1<x
\end{cases}
\]
This time the only missing cells are the isolated points $x=\pm 1, y=0$.  These would be included in a 3-layered sub-CAD, which is itself a complete CAD.  Hence in this case \texttt{LCAD} would give exactly the same output as \texttt{CADFull} (described in \cite{England13a}).

Note that each of \texttt{LCAD} command is distinct: re-running the projection phase and recomputing the higher dimensional cells.  \texttt{ProjectionCAD} also contains a recursive command to build layers incrementally.  It takes additional input as a list of cells already computed and sections that were previously discarded (in the final loop).  It then lifts over those discarded sections to produce cells of one dimension higher, as well as further discarded sections.  This recursive algorithm is described fully and verified in \cite{WBDE14}.

Algorithm \ref{alg:layeredCAD} allows for projection operators other than McCallum's.  In particular, it can be used with the TTICAD projection operator of \cite{BDEMW13} to build truth-table invariant sub-CADs, as for the following problem.
\begin{verbatim}
> f1 := x^2+y^2+z^2-1: g1:=x*y*z-1/4: 
> f2 := (x-1)^2+(y-1)^2+(z-1)^2-1: g2:=(x-1)*(y-1)*(z-1)-1/4:
> PHI := [ [f1,[g2]], [f2,[g2]] ]:
\end{verbatim}
The syntax here means that PHI defines two formulae: the first with EC defined by $f_1$ and additional constraints defined also by $g_1$; and the second with EC defined by $f_2$ and additional constraints defined by also $g_2$.  The TTICAD command could produce a decomposition sign-invariant for $f_1$ and $f_2$ and also for each $g_i$ where the corresponding $f_i$ is zero.
\begin{verbatim}
> TTICAD(PHI, [z,y,x]);
\end{verbatim}
\[
497
\]
The LTTICAD command can be used to build only the prescribed layers of cells:
\begin{verbatim}
> LTTICAD(PHI, 1, [z,y,x]), LTTICAD(PHI, 2, [z,y,x]),
  LTTICAD(PHI, 3, [z,y,x]), LTTICAD(PHI, 4, [z,y,x]);
\end{verbatim}
\begin{align*}
93, \, 299, \, 455, \, 497
\end{align*}
So for example, producing only the full-dimensional cells gives less than a fifth of the output.  It also takes less than half the computation time.

\section{Variety sub-CADs} 
\label{SEC:VCAD}

An {\bf equational constraint} (EC) is an equation, $f=0$, logically implied by the truth of a Tarski formula.  They may be explicit, like $f=0 \land \phi$, or implicit as $f_1f_2=0$ is in 
\[
(f_1 = 0 \land \phi_1) \lor (f_2 = 0 \land \phi_2).
\]  
The presence of an EC can reduce the number of projection polynomials required \cite[etc.]{McCallum1999b, BDEMW13} and \texttt{ProjectionCAD} already has the related commands \texttt{ECCAD} and \texttt{TTICAD} (described in \cite{England13b}).    Given a Tarski formula with EC $f=0$ we define a {\bf variety sub-CAD} as a truth-invariant {\scad} for the formula consisting only of cells lying in the variety defined by $f = 0$.
   
Algorithm \ref{alg:VCAD} describes how a variety sub-CAD may be produced.  We assume that all factors of the EC have the same main variable as the overall system ($x_n$) and so the EC is used only for the first projection and final lift.  Building variety sub-CADs outside of this restriction is considered in \cite{WBDE14} but not yet implemented in \texttt{ProjectionCAD}.

The projection proceeds as normal in step \ref{step:PV}, as does lifting to a CAD of $\R^{n-1}$ in step \ref{step:n-1}.  For the final lift we generate stacks only with respect to the EC in step \ref{step:lift}, noting that the projection theory used should ensure this is sufficient to conclude truth invariance (see \cite{BDEMW15} for details).  Further, only the sections in those stacks are retained for the output in step \ref{step:add} as only these describe the variety.  Since the underlying CAD algorithms that utilise ECs can return theoretical failure for input that is not well oriented we assume this is tested for during stack generation.  This approach was verified in \cite{WBDE14}.

Consider the formula $\varphi := (x^2+y^2-1 = 0) \land (xy-\tfrac{1}{4}>0) \land (x^3-y^2>0)$.  To identify the regions in $\mathbb{R}^3$ where this is true we could build a sign-invariant CAD for the three polynomials.  
\begin{verbatim}
CADFull([x^2+y^2-1,x*y-1/4,x^3-y^2], [x,y], method=McCallum): nops(%);
\end{verbatim}
\[
161
\]
However, since we see the formula has an EC we can use this to build a truth-invariant CAD for $\varphi$ with the \texttt{ECCAD} command:
\begin{verbatim}
ECCAD(  [x^2+y^2-1, [x*y-1/4,x^3-y^2]], [x,y]): nops(%);
\end{verbatim}
\[
73
\]
This more than halves the number of cells.  The \texttt{VCAD} command goes further and returns only those cells where the EC is satisfied (a sub-CAD) more than halving the output again:
\begin{verbatim}
VCAD(  [x^2+y^2-1, [x*y-1/4,x^3-y^2]], [x,y]): nops(%);
\end{verbatim}
\[
28
\]
Note that in all 3 cases it will be necessary to then test a sample point from each cell to see where the formula is true or false.  Although the cells outputted by VCAD guarantee the EC is satisfied, the two inequalities are only guaranteed truth-invariant, and so may still be false.

\begin{algorithm}[t]

\SetKwInOut{Input}{Input}\SetKwInOut{Output}{Output}

\Input{A choice of CAD projection \texttt{ProjOp}; input $F(\bm{x})$ (polynomials or formulae) in ordered variables $\bm{x}$; an EC $f=0$ (in which all factors have same mvar as $F$).
}

\Output{A (truth-invariant) variety {\scad} for $\varphi$, or FAIL if $F$ not well-oriented.}
\BlankLine

Set ${\bf P}$ to be the output from applying \texttt{ProjOp} to $F$\label{step:PV}\;

Compute a CAD of $\mathbb{R}^{n-1}$ for ${\bf P}$ and assign to $\mathcal{D}'$\label{step:n-1}\;

\If{ $\mathcal{D}'$ = {\rm FAIL}}{
\Return {\rm FAIL}\tcp*{${\bf P}$ is not well oriented}
}

Initialise $\mathcal{D} := []$\;

\For{$c \in \mathcal{D}'$}{
  Calculate $S := {\tt GenerateStack}(f, c)$\label{step:lift}\;
  \If{ $S$ = {\rm FAIL}}{
  \Return {\rm FAIL}\tcp*{$F$ is not well oriented}
  }
  \If{$|S| > 1$}{
    Add those cells in $S$ with last entry of their index even to $\mathcal{D}$\label{step:add}\;
  }
}

\Return $\mathcal{D}$\;

\caption{${\tt VCAD}(\varphi,f,{\bf x})$: Algorithm to produce variety {\scad}s.}
\label{alg:VCAD}
\end{algorithm}

\section{Additional functionality in \texttt{ProjectionCAD} version 3} 
\label{SEC:Others}

\subsubsection*{Combining sub-CADs}

We can build sub-CADs which are both variety and layered using the \texttt{LVCAD} command.  In fact, we can combine all the \texttt{ProjectionCAD} theory to build layered variety truth-table invariant sub-CADs with the \texttt{LVTTICAD} command.  More examples and details are given in \cite{WBDE14}, while in \cite{WDEB13} such a combination was used to solve a long-standing robot motion planning problem.

\subsubsection*{Cell distribution in CADs}

Tools are provided to compare the growth in CADs by layer, that is, the distributions of cells in CADs by dimensions.  Analysis in \cite{WEBD14} suggests a common distribution with little variation for the underlying problem, meaning layered sub-CADs can be used as heuristics to guide the construction of complete CADs. 

\subsubsection*{Using sub-CADs to avoid theoretical failure}  

Algorithms based on McCallum's theory can only run on input that is \textbf{well-oriented} (a definition particular to each individual operator, see for example \cite{McCallum1999b, BDEMW13}).  These conditions are checked during lifting and if not satisfied \texttt{ProjectionCAD} gives a warning.  For example:
\begin{verbatim}
> f := a*e+b*d+c*e+d+e: vars:=[a,b,c,d,e]:
> CADFull( [f], vars, method=McCallum  ): 
Warning, The input is not well-oriented (nullification on cell [1, 2, 2]).  
The output cannot be guaranteed correct.
\end{verbatim}
We can build a 1- or 2-layered sub-CAD for the polynomial without triggering the warning:
\begin{verbatim}
> LCAD([f],1,[a,b,c,d,e], method=McCallum,failure=err): 
> LCAD([f],2,[a,b,c,d,e], method=McCallum,failure=err): nops(%%), nops(%);
\end{verbatim}
\[
48, 148
\]
Hence we can get descriptions of not only the full dimensional cells but also the next dimension down without running into well-orientedness issues.  

\section{Summary}
\label{SEC:Summary}

We have described the new functionality present in the third release of \texttt{ProjectionCAD}, which focusses on algorithms to build sub-CADs.  The underlying theory is discussed further in \cite{WDEB13, WEBD14, WBDE14} and the present report should be accompanied by the code and an introductory \textsc{Maple} worksheet demonstrating the full functionality of the package.
A key topic for future work is building variety sub-CADs when the EC does not have all factors in the main variable.  Various approaches are laid out in Section 2.1 of \cite{WBDE14} and an analysis of these is ongoing.

\begin{footnotesize}
\bibliography{CAD}{}
\bibliographystyle{plain}
\end{footnotesize}

\end{document}